\nonstopmode
\documentclass{jpconf}

\usepackage{amsmath}
\usepackage{iopams}
\usepackage{graphicx}
\graphicspath{{./}{../graphics/}}
\usepackage{hyperref}

\newcommand{\Z}{\mathbb{Z}}
\newcommand{\dif}{\,{\rm d}}
\DeclareMathOperator{\vol}{vol}

\newcommand{\arXiv}[2]{\href{http://arxiv.org/pdf/#1}
  {\emph{Preprint} #1#2}}

\begin{document}

\title{Some physical consequences of an exact vacua distribution in
  the Bousso-Polchinski Landscape}

\author{C\'esar Asensio and Antonio Segu\'{\i}}

\address{Departamento de F\'{\i}sica Te\'orica, Universidad de
  Zaragoza, Spain}

\ead{casencha@unizar.es}

\begin{abstract}
  The Bousso-Polchinski (BP) Landscape is a proposal for solving the
  Cosmological Constant Problem.  The solution requires counting the
  states in a very thin shell in flux space.  We find an exact formula
  for this counting problem which has two simple asymptotic regimes,
  one of them being the method of counting low $\Lambda$ states given
  originally by Bousso and Polchinski.  We finally give some
  applications of the extended formula: a robust property of the
  Landscape which can be identified with an effective occupation
  number, an estimator for the minimum cosmological constant and a
  possible influence on the KKLT stabilization mechanism.
\end{abstract}

\section{Introduction}
\label{sec:intro}

The \emph{cosmological constant problem} \cite{WW2,B-CC} is the
smallness of the observed value\footnote{We use reduced Planck units
  in which $8\pi G = \hbar = c = 1$.} $\Lambda_{\text{obs}} =
1.5\times10^{-123}$ \cite{SN-1,SN-2} when compared with naive
expectations from particle physics.  An attempt for a solution is
proposed in the Bousso-Polchinski Landscape \cite{BP}, in which a
large amount $J$ of quantized fluxes of charges
$\{q_j\}_{i=1,\cdots,J}$ leads to an effective cosmological constant
\begin{equation}
  \label{eq:2}
  \Lambda = \Lambda_0 + \frac{1}{2}\sum^J_{j=1}n_j^2 q_j^2\,.
\end{equation}
In (\ref{eq:2}), $\Lambda_0$ is a negative number of order $-1$, and
the integer $J$-tuple $(n_1,\cdots,n_J)$ characterizes each of the
vacua of the Landscape, which is a finite subset (yet an enormous one)
of an infinite lattice comprising the nodes with cosmological constant
smaller than some value $\Lambda_1=\mathcal{O}(1)$.  For large $J$ and
incommensurate charges $\{q_j\}$ this model contains states of small
$\Lambda$.  The problem arises now as how to count them.

Each state in the Bousso-Polchinski Landscape can be viewed as a node
of a lattice in flux space surrounded by a cell of volume $\vol
Q=\prod^J_{i=1}q_i$.  On the other hand, each value of the
cosmological constant $\Lambda_0 \le \Lambda \le \Lambda_1$ defines a
ball $\mathcal{B}^J(r)$ in flux space of radius $r =
\sqrt{2(\Lambda-\Lambda_0)}$.  The BP counting argument
\cite{BP,Shenker} consists of computing the number of states inside a
ball of radius $r$ by
\begin{equation}
  \label{eq:7}
  \Omega_J(r) = \frac{\vol\mathcal{B}^J(r)}{\vol Q}\,.
\end{equation}

Nevertheless, as the authors of \cite{BP} point out, this argument is
not valid when any of the charges $q_i$ exceed $R_0/\sqrt{J}$.
Therefore, we will propose an exact counting formula which is reduced
to (\ref{eq:7}) in the appropiate regime.

\section{The BP Landscape degeneracy}
\label{sec:omega}

\subsection{The exact representation}
\label{sec:exact}

We start with the number of nodes in the lattice inside a sphere in
flux space of radius $r$.  This magnitude is called $\Omega_J(r)$
above.  It can be given in terms of the characteristic function of an
interval $I$, $\chi_I(t) = 1$ if $t\in I$ and 0 otherwise:
\begin{equation}
  \label{eq:15}
  \Omega_J(r) = \sum_{\lambda\in\mathcal{L}} \chi_{[0,r]}(\|\lambda\|)
  \,.
\end{equation}
Expression (\ref{eq:15}) is exact and finite, and it is equivalent to
directly counting the nodes (the ``brute-force'' counting method),
hence it cannot be used in order to obtain numbers as in (\ref{eq:7}).

The density of states associated to (\ref{eq:15}) is $\omega_J(r) =
\frac{\partial \Omega_J(r)}{\partial r} =
2r\sum_{\lambda\in\mathcal{L}} \delta\bigl(r^2-\|\lambda\|^2\bigr)$
which will be called the ``BP Landscape degeneracy''.  The counting
function $\Omega_J(r)$ is a stepwise monotonically non-decreasing
function, and thus its derivative $\omega_J(r)$ is a sum of Dirac
deltas.  We can express these Dirac deltas as integrals in complex
plane along a vertical line $\gamma$ crossing the positive real axis
in complex plane.  We obtain
\begin{equation}
  \label{eq:23}
    \omega_J(r) 
    = \frac{2r}{2\pi i}
    \int_\gamma  e^{sr^2}
    \Biggl[\prod_{j=1}^J
    \vartheta(sq_j^2)\Biggr] \dif s
    \,.
\end{equation}
The sum is hidden in the function $\vartheta(s) = \sum_{n\in\Z}
e^{-sn^2}$, valid for $\mathrm{Re}\,s>0$.

The integration of (\ref{eq:23}) with the initial condition
$\Omega_J(0)=1$ gives $\Omega_J(r)$.

\subsection{The large distance (or BP) regime}
\label{sec:BP-regime}

Now we will turn to the approximate evaluation of $\omega_J(r)$.  For
this purpose we need the asymptotic behavior of $\vartheta$ function.
There is a middle regime where the asymptotic regimes are not accurate
enough, and we have computed numerically all the quantities.

The first case is $s\to0$.  In this regime, we make the integration
contour pass near the origin, where $\vartheta$ has a singularity.
Assuming that the main contribution to the integral will come from
this region, we can replace $\vartheta$ by its asymptotic value when
$s\to0$ and we obtain
\begin{equation}
  \label{eq:34}
  \omega_J(r) \approx \frac{2r}{2\pi i}\int_\gamma e^{sr^2}
  \Biggl[\prod_{j=1}^J \sqrt{\frac{\pi}{q_j^2 s}} \Biggr] \dif s
  = \frac{\pi^{\frac{J}{2}}}{\vol Q}\,\frac{2r}{2\pi i} \int_\gamma
  e^{sr^2}\frac{\dif s}{s^{\frac{J}{2}}} 
  = \frac{2\pi^{\frac{J}{2}}}{\Gamma(\frac{J}{2})}\,
  \frac{r^{J-1}}{\vol Q}
  \,.
\end{equation}
Equation \eqref{eq:34} is the derivative of (\ref{eq:7}), that is, BP
count.  It is valid for large $r$ distances, $h = \frac{Jq^2}{r^2} <
\frac{2}{e} \approx 0.736$.

\subsection{The small distance regime}
\label{sec:small-r}

In this case we are in the regime in which the asymptotic expansion of
$\vartheta$ for large values of its argument is valid.  We can
estimate the integral using the saddle point approximation.
Unfortunately, we cannot solve the saddle point equation in closed
form for arbitrary charges.  Nevertheless, in the simplest case in
which all charges are equal $q_1=\cdots=q_J=q$, we obtain
\begin{equation}
  \label{eq:47}
  \omega_J(r) = \frac{(2h-2)^{\frac{J}{h}}}{q\sqrt{2\pi h}}
  \Bigl(\frac{h}{h-1}\Bigr)^{J+\frac{1}{2}}\,.
\end{equation}
The saddle point and the asymptote are in the same region if $h =
\frac{Jq^2}{r^2} > 1 + \frac{e^2}{2} \approx 4.694$.

\section{Applications}
\label{sec:app}

\subsection{Number of states in the Weinberg Window}
\label{sec:NWW}

The number of states of positive cosmological constant bounded by a
small value $\Lambda_\varepsilon$ is the number of nodes of the
lattice in flux space whose distance to the origin lies in the
interval $[R,R_{\varepsilon}]$, where $R = \sqrt{2|\Lambda_0|}$
and $R_\varepsilon = \sqrt{2(\Lambda_\varepsilon - \Lambda_0)} \approx
R + \frac{\Lambda_\varepsilon}{R}$ so that the width of the shell
is $\varepsilon = \frac{\Lambda_\varepsilon}{R}$.  If
$\Lambda_\varepsilon$ is the width of the anthropic range
$\Lambda_{\text{WW}}$ (the so-called Weinberg Window), then the number
of states in it is
\begin{equation}
  \label{eq:55}
  \mathcal{N}_{\text{WW}} = \frac{\omega_J(R)}{R}\Lambda_{\text{WW}}\,.
\end{equation}
Computation of $\omega_J(R)$ should be done along the lines of the
previous section.

\subsection{Typical number of non-vanishing fluxes }
\label{sec:alpha-star}

The set of nodes inside a thin shell of width
$\varepsilon=R_\varepsilon-R$ above radius $R$ will be called
$\Sigma_\varepsilon$.  We will assume that $\varepsilon$ is smaller
than the charges $q_i$ so that (\ref{eq:55}) is valid but
$\mathcal{N}_\varepsilon\gg1$.

We find that the typical number of non-vanishing components of a state
drawn randomly from $\Sigma_\varepsilon$ is $J$ for the cases $J=2,3$.
We wonder whether it happens for all $J$.  We will answer this
question by computing the fraction of states in the shell having a
fixed fraction $\alpha$ of non-vanishing components.

\begin{figure}[htbp]
  \centering
  \includegraphics[width=0.5\textwidth]{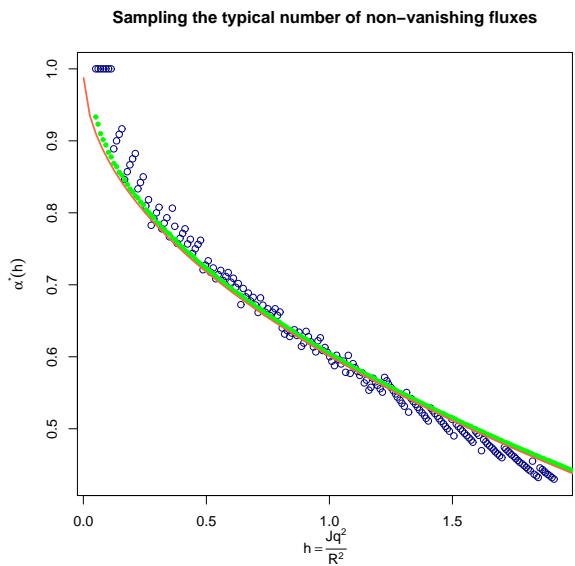}
  \quad
  \begin{minipage}[b]{0.4\linewidth}
    \caption{\label{fig:alpha-star} Samples of the typical number of
      non-vanishing fluxes.  Two sampling methods have been used.  The
      saddle point solution is also shown (red line).}
  \end{minipage}
\end{figure}

When a state $\lambda$ is selected at random from $\Sigma_\varepsilon$
with uniform probability, $\alpha$ becomes a discrete random variable
taking values in the $[0,1]$ interval.  Assuming equal charges, its
probability distribution is given by (see \cite{ESTE} for details)
\begin{equation}
  \label{eq:58}
  P(\alpha) = \frac{2R}{\omega_J(R)}\binom{J}{\alpha J}
  \frac{1}{2\pi i}\int_\gamma e^{\phi(s,\alpha)}\dif s
  \quad\text{with}\quad
  \phi(s,\alpha) = sR^2 + \alpha J\log\bigl[\vartheta(q^2s) - 1\bigr]
  \,.
\end{equation}
It can be seen that $P(\alpha)$ is locally Gaussian around its peak
$\alpha^*(h)$, with standard deviation $\sim1/\sqrt{J}$.
$\alpha^*(h)$ is the typical number of non-vanishing fluxes in the
shell $\Sigma_\epsilon$ (and essentially also in the whole Landscape).
Its computation must be done numerically, either using the saddle
point method on \eqref{eq:58}, or by statistical sampling, see
\cite{ESTE}.  The results are plotted in figure \ref{fig:alpha-star}.

\subsection{Estimating the minimum positive cosmological constant}
\label{sec:lambda_min}

We can roughly estimate the explicit dependence of the minimum
positive cosmological constant with respect to the parameters of the
Landscape.  We will call $\Lambda^*$ the actual minimum value, and
$\Lambda_\varepsilon$ the corresponding estimator.

In the case of incommensurate charges, the symmetry degeneracy of a
state is $2^{J\alpha^*}$, so that we have $\Lambda_\varepsilon \approx
\frac{2^{J\alpha^*}R}{\omega_J(R)}$.  We can check this estimate with
brute-force data for low $J$ and we find a good agreement in the
statistical sense.

\subsection{A possible influence on the KKLT mechanism}
\label{sec:KKLT}

The Giddings-Kachru-Polchinski model \cite{GKP} is a more realistic
approach to the true string theory Landscape, and it can be endowed
with a mechanism for fixing the compactification moduli, the so-called
KKLT mechanism \cite{KKLT}.  In this model, moduli are stabilized by
the presence of fluxes and corrections to the superpotential coming
from localized branes.

As far as we know, there is no combination between the BP Landscape
and the KKLT mechanism, in the sense that there is no known realistic
model in which all moduli are fixed and a large amount of three-cycles
lead to an anthropic value of the cosmological constant.  Let us
assume that such a model will be built in the near future.  If the
$\alpha^*(h)$ curve discussed in the section \ref{sec:alpha-star} can
be generalized, that is, the typical occupation number of the fluxes
is different from 1, then there will be a finite fraction $1-\alpha^*$
of three-cycles with vanishing flux.  This fraction of vanishing
fluxes can spoil the stabilization mechanism.

\section{Conclusions}
\label{sec:conc}

We have developed an exact formula for counting states in the
Bousso-Polchinski Landscape which reduces to the volume-counting one
in certain (BP) regime.  Numeric computations and brute-force searches
have been carried out to check the results of our analytic
approximations, and we have found remarkable agreement in all explored
regimes.

In particular, we have discovered a robust property of the BP
Landscape, the typical fraction of non-vanishing fluxes $\alpha^*(h)$,
which reveals the structure of the lattice inside a sphere for large
$J$ as the union of hyperplane portions of effective dimension near
$J\alpha^*$.  This result is important in computing degeneracies,
which are used in estimating the minimum cosmological constant, and it
could be an obstacle for a realistic implementation of the KKTL moduli
stabilization mechanism.

\ack

We would like to thank Pablo Diaz, Concha Orna and Laura Segui for
carefully reading this manuscript.  We also thank Jaume Garriga for
useful discussions and encouragement.  This work has been supported by
CICYT (grant FPA-2006-02315 and grant FPA-2009-09638) and DGIID-DGA
(grant 2007-E24/2). We thank also the support by grant A9335/07 and
A9335/10 (F\'{\i}sica de alta energ\'{\i}a: Part\'{\i}culas, cuerdas y
cosmolog\'{\i}a).

\section*{References}
\label{sec:refs}


\begin{thebibliography}{99}


\bibitem{WW2} Weinberg S 1989
  \emph{Rev. Mod. Phys.} \textbf{61}, 1-23
  
\bibitem{B-CC} Bousso R 2008
  \emph{Gen. Rel. Grav.} \textbf{40} 607
  (\arXiv{0708.4231}{ [hep-th]})


\bibitem{SN-1} Perlmutter S \emph{et al.} 1999
  \emph{Astrophys. J.} \textbf{517} 565-586
  (\arXiv{astro-ph/9812133}{})

\bibitem{SN-2} Riess A G \emph{et al.} 1998
  \emph{Astron. J.} \textbf{116} 1009-1038
  (\arXiv{astro-ph/9805201}{})


\bibitem{BP} Bousso R and Polchinski J 2000
  \emph{JHEP} \textbf{06} 006 (\arXiv{hep-th/0004134}{})

\bibitem{Shenker} Clifton T, Shenker S and Sivanandam N 2007
  \emph{JHEP} \textbf{0709} 034 (\arXiv{0706.3201}{ [hep-th]})


\bibitem{ESTE} Asensio C and Segui A 2010
  \arXiv{1003.6011}{ [hep-th]}


\bibitem{GKP} Giddings S B, Kachru S and Polchinski J 2002
  \emph{Phys. Rev.} D \textbf{66} 106006
  (\arXiv{hep-th/0105097}{})

\bibitem{KKLT} Kachru S, Kallosh R, Linde A and Trivedi S 2003
  \emph{Phys. Rev.} D \textbf{68} 046005
  (\arXiv{hep-th/0301240}{})

\end{thebibliography}
\end{document}